\documentclass[%
 reprint,
 amsmath,amssymb,
 aps,
]{revtex4-2}

\usepackage{graphicx}
\usepackage{dcolumn}
\usepackage{bm}
\usepackage{hyperref}

\begin{document}

\preprint{APS/123-QED}

\title{Rabi Oscillation of High Partial Wave Interacting Atoms in Deep Optical Lattice}

\author{Zeqing Wang}

\author{Ran Qi}%
\email{qiran@ruc.edu.cn}
\affiliation{%
Department of Physics, Renmin University of China, Beijing, 100872, P. R. China
}%

\date{\today}

\begin{abstract}
  Motivated by the recent experiment of p-wave interacting $^{40}\mathrm{K}$ atom gases in a deep optical lattice, we investigated the Rabi oscillation for any partial wave interacting quantum gases in a deep optical lattice. We first review the solution for two particles interacting by any partial wave in a harmonic trap by using pseudopotential.
  We generalize the model used in recent work to any partial wave cases,
  repeated some of the theoretic and experiment results for p-wave interacting $^{40}\mathrm{K}$ atoms in a deep optical lattice
  and show the results for the d-wave case.
  Our results may be useful for future experiments and may stimulate further theoretic investigation.
\end{abstract}

\maketitle


\section{Introduction}

High partial wave interacting many-body systems has attracted much interest in recent years due to the existence of novel states such as topological superfluids and Majorana zero modes~\cite{PhysRevLett.86.268, doi:10.7566/JPSJ.85.022001, Alicea_2012}.
The experiments are also under rapid development.
For example, Feshbach resonance enhanced interacting ultracold atomics gas near the p-wave, d-wave and g-wave resonance have been realized in recent years~\cite{luciuk_evidence_2016, PhysRevLett.119.203402, PhysRevA.98.022704, PhysRevA.98.020702, PhysRevA.100.050701, yao_degenerate_2019, PhysRevLett.125.263402, PhysRevResearch.3.033269, zhang_transition_2021, PhysRevA.106.023322, PhysRevA.107.053322, Venu2023, Shi_2023}.
However, there are three-body loss problems in high partial wave interacting quantum gases.
In a recent experiment, p-wave interacting $^{40}\mathrm{K}$ atoms are loaded into an optical lattice~\cite{Venu2023}. When the optical lattice is deep enough, there are only two atoms in each site. Therefore there is no three-body loss.

Theoretically, such a deep optical lattice system can be described by a pseudopotential interacting two-body problem in a harmonic trap.
The ultracold atomic gases are dilute, this implies that the typical distance $d\sim n^{1/3}$ with $n$ the density of the atomic gases is much larger than the effective range $r_0$ of the interaction between atoms, i.e. $d\gg r_0$. Therefore we will only consider the $r \gg r_0$ long-range physics, where $r$ is the distance between two atoms.


In the $r\gg r_0$ long-range, the interaction strength between atoms is described by a momentum-dependent scattering phase shift $\delta_l(k)$ with $l$ the partial wave number and $k$ the relative momentum for two atoms.
The phase shift $\delta_l(k)$ is decided by the short-range details of the scattering potential between atoms, which is usually unknown and complicated. Fortunately, we only care about the long-range physics in ultracold quantum gases, so we can use a more simple potential, i.e. a pseudopotential, to describe the interaction between atoms as long as such a potential gives the same phase shift.

Pseudopotential for any partial wave was derived and applied to many-body problems by Kerson Huang and C. N. Yang in 1957~\cite{PhysRev.105.767}.
There are some typos in Kerson Huang and C. N. Yang's results and have been corrected in~\cite{PhysRevA.72.044701, PhysRevLett.94.023202}.
In the derivation in~\cite{PhysRevLett.94.023202},
the pseudopotential was assumed  as a hard spherical shell potential with radius $s$,
\begin{align}\label{eq:pseudopotential}
    V(r) = \sum_l V_l(r)= \frac{\hbar^2}{2m}\sum_l\lim_{s\to 0}\delta(r - s)\hat{O}_l(r).
  \end{align}
The operator $\hat{\mathcal{O}}_l(r)$ for $l$ partial wave can be derived from the boundary condition as
\begin{align}
    \hat{O}_l(r) = - \frac{(2l+1)!!}{(2l)!!}
    \frac{\tan \delta_l(k)}{k^{2l+1}} \frac{1}{s^{l+2}}
    \frac{\partial^{2l+1}}{\partial r^{2l+1}} r^{l+1}.
  \end{align}
This pseudopotential will give the scattering phase shift $\delta_l(k)$. Therefore we can use this pseudopotential as scattering potential in theoretic study as long as we are only concerned with long-range physics.



In Sec~\ref{subsec:two-body}, we first review the solution for two particles in a harmonic trap given in~\cite{PhysRevLett.94.023202}.
Then we review and generalize the p-wave theoretic model in~\cite{Venu2023} to any partial wave in Sec~\ref{subsec:optical-lattice}.
Finally we show the results for the p-wave and d-wave cases in Sec~\ref{sec:results}. We summarize our work in Sec~\ref{sec:summary}.

\section{Model}

\subsection{Two-body relative motion in a harmonic potential}\label{subsec:two-body}

The trap for atoms in a very deep lattice can be approximately considered as harmonic trap theoretically~\cite{Venu2023}.
The two-body problem in a harmonic trap was investigated in~\cite{Busch1998} for s-wave pseudopotential interaction and in~\cite{PhysRevA.70.042709} for p-wave interaction. The authors give the analytic solution of the energy spectrum and the corresponding wave functions. For higher partial wave interacting cases, the results were given in~\cite{PhysRevLett.94.023202}.
We will first review the solution for two free particles in a harmonic trap. Then we review the results for any partial wave interacting cases.

\begin{figure*}[htb]
  \centering
  \includegraphics{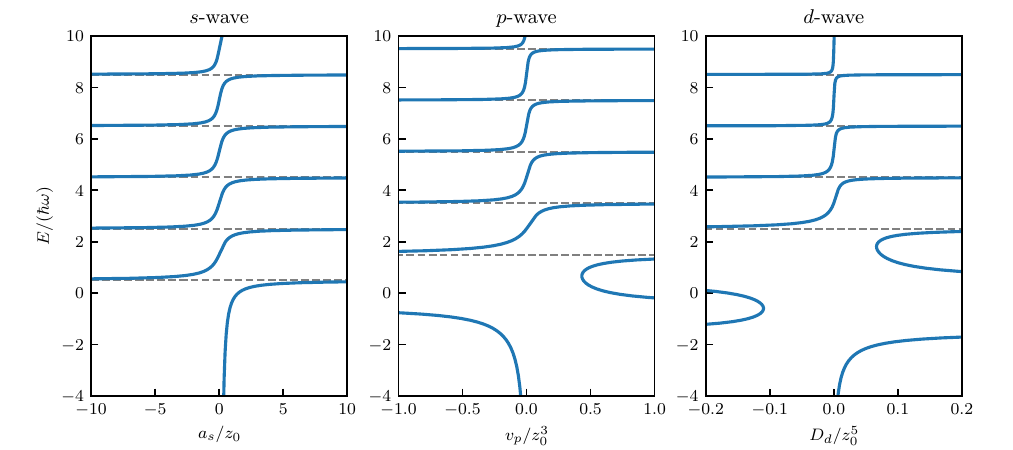}
  \caption{\label{fig:energy_spectrum}
  The energy spectrum for two (a)~s-wave, (b)~p-wave, (c)~d-wave interacting particles in a harmonic trap. The results for s-wave and p-wave are consistent with the results in~\cite{Busch1998, PhysRevA.70.042709}.
  }
\end{figure*}

First, we consider two free particles with mass $m_1$ and $m_2$ in a harmonic potential with frequency $\omega$. The Hamiltonian is
  \begin{align}
    H_0 = - \frac{\hbar^2}{2m_1}\nabla_1^2 - \frac{\hbar^2}{2m_2}\nabla_2^2
    + \frac{1}{2}\omega^2\left( m_1 r_1^2 + m_2 r_2^2 \right),
  \end{align}
where $r_1 = |\mathbf{r_1}|$, $r_2 = |\mathbf{r_2}|$, $\mathbf{r}_1$ and $\mathbf{r}_2$ are the spacial coordinates of two particles.
Because both the kinetic energy and harmonic trap potential are in quadratic form, we can separate the Hamiltonian into the center of mass motion and relative motion. The center of mass motion Hamiltonian is
  \begin{align}
    H_{\mathrm{CM}} = - \frac{\hbar^2}{2M}\nabla_R^2
    + \frac{1}{2} M \omega^2 R^2,
  \end{align}
where $M = m_1 + m_2$ is the total mass, $\mathbf{R} = (m_1 \mathbf{r}_1 + m_1 \mathbf{r}_1)/(m_1 + m_2)$ is the center of mass coordinate. The relative motion Hamiltonian is,
  \begin{align}\label{eq:H-rel-free}
    H_{\mathrm{rel},0} = - \frac{\hbar^2}{2\mu}\nabla_r^2
    + \frac{1}{2} \mu \omega^2 r^2,
  \end{align}
where $r = |\mathbf{r}|$, $\mathbf{r} = \mathbf{r}_1 - \mathbf{r}_2$ is the relative coordinate,
$\mu = m_1 m_2 /(m_1 + m_2)$ is the reduced mass.
The center of mass motion Hamiltonian $H_{\mathrm{CM}}$ and relative motion Hamiltonian $H_{\mathrm{rel},0}$ are in the same form. Their solutions are both the well known three-dimensional isotropic harmonic oscillator solutions. We write the relative motion solution here.
The wave function for relative motion in spherical coordinates is
  \begin{align}
    \label{eq:ho-wf-free}
    \Psi_{\nu, l, m} = \psi_{\nu, l} (r) Y_l^m(\hat{\mathbf{r}}),
  \end{align}
 where $\hat{\mathbf{r}} = \mathbf{r}/r$ is the direction of the relative position vector $\mathbf{r}$, $Y_l^m(\hat{\mathbf{r}})$ is the spherical harmonic function.
 The states are described by three quantum numbers. Quantum number $\nu$ is a  non-negative integer, $l$ and $m$ is the angular momentum quantum number and the porjection in $z$ direction.
 The radical wave function $\psi_{\nu, l} (r)$ is
  \begin{align}
    \label{eq:radical-wf-free}
    \psi_{\nu, l} (r) =
     \mathcal{A}_{\nu, l} \left( \frac{r}{z_0} \right)^l e^{-\frac{r^2}{2 z_0^2}} \mathcal{L}_{\nu}^{l+\frac{1}{2}} \left(\frac{r^2}{z_0^2}\right) ,
  \end{align}
where $\mathcal{A}_{\nu, l}$ is a normalization factor,
  \begin{align}
    \mathcal{A}_{\nu, l} = \frac{1}{z_0^{3/2}}
    \sqrt{\frac{1}{\sqrt{\pi}}\frac{2^{\nu + l + 2} \nu!}{(2\nu + 2l + 1)!!}},
  \end{align}
with $z_0 = \sqrt{\hbar/(\mu\omega)}$ is the characteristic length of the harmonic trap. $\mathcal{L}$ is the generalized Laguerre polynomials.
The energy eigenvalues for wave function $\Psi_{\nu, l, m}$ is
  \begin{align}
    \label{eq:ho-energies}
    E_{\nu, l} = \left( 2 \nu + l + \frac{3}{2} \right)\hbar\omega,
  \end{align}
which only depends on the quantum numbers $\nu$ and $l$.

Next, we consider two interacting particles in a harmonic trap.
The interaction between the two particles is described by scattering phase shift is $\delta_l(k)$ for long-range physics.
Therefore we can use the pseudopotential to describe the scattering potential between these two particles.
The center of mass motion is the same as the non-interacting atoms case.
The relative motion Hamiltonian can be written as,
  \begin{widetext}
  \begin{align}
    H_{\mathrm{rel}} = - \frac{\hbar^2}{2\mu}\nabla_r^2
     + \frac{1}{2} \mu \omega^2 r^2
    -\frac{\hbar^2}{2\mu}
    \lim_{s\to 0}\delta(r - s)
    \frac{(2l+1)!!}{(2l)!!}
    \frac{\tan \delta_l(k)}{k^{2l+1}} \frac{1}{s^{l+2}}
    \frac{\partial^{2l+1}}{\partial r^{2l+1}} r^{l+1},
  \end{align}
\end{widetext}
where $k$ is the relative momentum.
The energy eigenvalues $E^{\mathrm{pseudo}}$ satisfy the equation~\cite{PhysRevLett.94.023202},
  \begin{align}
    \label{eq:pseudo-3D-harmonic-energy}
    \frac{\pi}{2}\frac{(-1)^l \left[(2l+1)!!\right]^2}{\left[ \Gamma(l+3/2) \right]}
    \frac{\Gamma(-\nu)}{\Gamma(-\nu - l - 1/2)}
    = \frac{1}{a_l^{2l+1}(k)}
  \end{align}
where $\Gamma$ is the gamma function, $\nu = E^{\mathrm{pseudo}}/2 - l/2 - 3/4$.
The generalized scattering length for $l$ partial wave is,
  \begin{align}
    a_l^{2l+1}(k) = - \frac{\tan\delta_l(k)}{k^{2l+1}}.
  \end{align}
In experiments, the generalized scattering length $a_l(k)$ can be tuned by magnetic Feshbach resonance. According to Eq.~\eqref{eq:pseudo-3D-harmonic-energy}, for each interacting strength, there will be a series energy eigenvalues. The energy eigenvalues are labeled by quantum numbers $\nu$ and $l$. When the interaction strength adiabatically tuned to zero, the energy eigenvalues will become the energy given by Eq.~\eqref{eq:ho-energies}.

We show the energy spectrum for the s-wave, p-wave, and d-wave cases in Fig.~\ref{fig:energy_spectrum}. Here, for low energy scattering, we only consider the leading order in generalized scattering length $a_l(k)$ of $k$.
For s-wave, $a_l^{2l+1}(k) \approx a_s$, is the scattering length.
For p-wave, $a_l^{2l+1}(k)\approx v_p$, is the scattering volume.
For d-wave $a_l^{2l+1}(k) \approx D_d$, is the super volume.
The results in Fig~\ref{fig:energy_spectrum}~(a) for s-wave and Fig~\ref{fig:energy_spectrum}~(b) for p-wave are consistent with the results in~\cite{Busch1998, PhysRevA.70.042709}.

When $r>s$, the corresponding wave function is~\cite{PhysRevLett.94.023202}
\begin{align}
  \label{eq:ho-wf-pseudo}
  \Psi^{\mathrm{pseudo}}_{\nu, l, m}(r) = \psi^{\mathrm{pseudo}}_{\nu,l} Y_l^m(\hat{\mathbf{r}}),
\end{align}
where the radical wave function is
\begin{align}
  \label{eq:radical-wf-pseudo}
  \psi^{\mathrm{pseudo}}_{\nu, l}(r) = \mathcal{A}_{\nu, l}^{\mathrm{pseudo}} \left( \frac{r}{z_0} \right)^{l}
  e^{- \frac{r^2}{2z_0^2}} U\left(-\nu, l + \frac{3}{2}, \frac{r^2}{z_0^2}\right),
\end{align}
with $U$ the Tricomi's confluent hypergeometric function. It is worth noting that we should choose a finite cutoff when we numerically calculate the normalization factor $\mathcal{A}_{\nu, l}^{\mathrm{pseudo}}$ because of its singularity in the origin.

\subsection{Deep Optical Lattice}\label{subsec:optical-lattice}

In this section, we generalize the model in the recent p-wave experiment~\cite{Venu2023} to any partial wave interaction using the solution of two-body problem in a harmonic trap given in Sec~\ref{subsec:two-body}.

We consider two-component atomics gases, we will call these two states spin and label as ${|\uparrow\rangle}$ and ${|\downarrow\rangle}$. They are the hyperfine states of atoms in experiments. The Zemann energy difference between these two states is $\hbar \omega_{\mathrm{zs}}$. There only exist interactions between two ${|\uparrow\rangle}$ particles due to Feshbach resonance.

The Hamiltonian for one particle in basis $\{ {|\uparrow\rangle}, {|\downarrow\rangle} \}$ is
\begin{align}
  \frac{\hbar \omega_{\mathrm{zs}}}{2}
  \begin{pmatrix}
    1 & 0 \\
    0 & -1
  \end{pmatrix}.
\end{align}
Then we couple these two states by an RF pulse with frequency $\omega_{\mathrm{rf}}$ and Rabi frequency $\Omega_1$. In a $e^{-\mathrm{i} \omega_{\mathrm{rf}}\hat{\sigma}_z t}$ rotated frame, the Hamiltonian for single particle can be written as,
\begin{align}
  H_{\mathrm{sing}} = \frac{\hbar}{2}
  \begin{pmatrix}
    \delta & \Omega_1 \\
    \Omega_1 & -\delta
  \end{pmatrix},
\end{align}
where $\delta = \omega_{\mathrm{zs}} - \omega_{\mathrm{rf}}$ is the detuning between the Zemann energy difference and the RF pulse frequency.

Without considering the spacial motion, the spin degree of freedom Hamiltonian of two particles in the direct product basis $\{ {|\uparrow\uparrow\rangle}, {|\uparrow\downarrow\rangle}, {|\downarrow\uparrow\rangle}, {|\downarrow\downarrow\rangle}\}$ is the sum of two single-particle Hamiltonians,
\begin{align}
  H_{\mathrm{spin}} =& H_{\mathrm{sing}} \otimes I_2 + I_2 \otimes H_{\mathrm{sing}} \\
  =& \frac{\hbar}{2}\begin{pmatrix}
    2\delta  & \Omega_1 & \Omega_1 & 0 \\
    \Omega_1 & 0 & 0 & \Omega_1 \\
    \Omega_1 & 0 & 0 & \Omega_1 \\
    0 & \Omega_1 & \Omega_1 & -2\delta
  \end{pmatrix},
\end{align}
where $I_2$ is the $2\times 2$ identity matrix.
Now, we shall change to the more convenient coupled basis, which can be divided into two groups according to the exchange symmetry of two particles. One group is the triplets,
\begin{align}
  |T, \uparrow\uparrow \rangle =& |\uparrow\uparrow\rangle , \\
  |T, \uparrow\downarrow \rangle =& \frac{1}{\sqrt{2}}\left(
    |\uparrow\downarrow\rangle + |\downarrow\uparrow\rangle
   \right), \\
   |T, \downarrow\downarrow \rangle =& |\downarrow\downarrow\rangle .
\end{align}
The other group is singlet,
\begin{align}
  |S, \uparrow\downarrow \rangle = \frac{1}{\sqrt{2}}\left(
    |\uparrow\downarrow\rangle - |\downarrow\uparrow\rangle
   \right).
\end{align}
We will only consider the triplets because there is no coupling between the triplets and the singlet.

Now we consider the relative spacial motion of these two particles. For two particles in state ${|T, \uparrow\downarrow \rangle}$ and ${|T, \downarrow\downarrow \rangle}$, the spacial motion are described by the relative motion of two non-interacting particles in a harmonic trap whose Hamiltonian is Eq.~\eqref{eq:H-rel-free} $H_{\mathrm{rel}, 0}$.
While for two particles in state ${|T, \uparrow\uparrow \rangle}$, there exists interaction described by pseudopotential. Therefore we can write the spacial motion Hamiltonian as
\begin{align}
  \hat{H}_{\mathrm{rel},0 } + \hat{V},
\end{align}
where $\hat{V}$ is the spin-dependent interaction,
\begin{align}
  \hat{V} = |T, \uparrow\uparrow \rangle \langle T, \uparrow\uparrow | V_l(r),
\end{align}
with $V_l(r)$ the pseudopotential in Eq.~\eqref{eq:pseudopotential}.
The spacial wave function for states $|T, \uparrow\uparrow \rangle$ is $\Psi^{\mathrm{pseudo}}_{\nu, l, m} = \langle \mathbf{r} | \nu, l, m\rangle_{\mathrm{pseudo}}$ give by Eq.~\eqref{eq:ho-wf-pseudo}, and $\Psi_{\nu, l, m} = \langle \mathbf{r} | \nu, l, m\rangle$  give by Eq.~\eqref{eq:ho-wf-free} for other two states.

\begin{figure}[htb]
  \centering
  \includegraphics{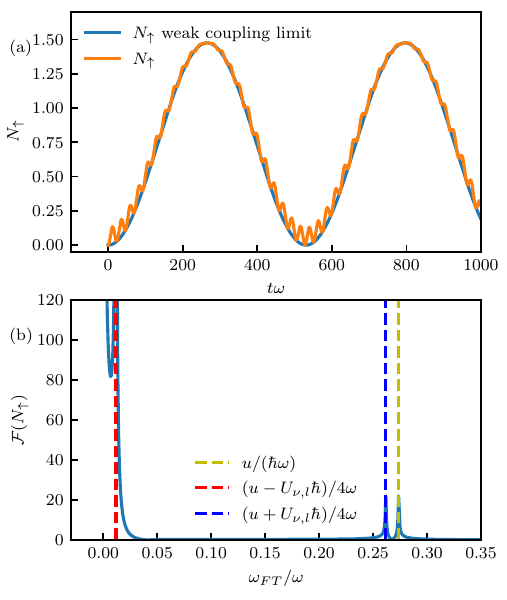}
  \caption{\label{fig:rabi_osc}
    (a) p-wave Rabi oscillation for $^{40}\mathrm{K}$ atoms. (b) The Fourier transform of the pair oscillation. The RF pulse frequency is in resonance with the eigen state with energy $E = 3 \hbar\omega$. The trap frequency is $\omega = 2\pi \times 129~\mathrm{kHz}$.The Rabi frequency of the RF pulse is $\Omega_1=2\pi\times 8.83~\mathrm{kHz}$. The cutoff for $\eta$ is $50~a_0$.
    }
\end{figure}

We choose the following three states as a set of basis,
\begin{align}\label{eq:basis}
  &|T, \uparrow\uparrow \rangle \otimes |\nu, l, m\rangle_{\mathrm{pseudo}},\\ \nonumber
  &|T, \uparrow\downarrow \rangle  \otimes |\nu, l, m\rangle, \\ \nonumber
  &|T, \downarrow\downarrow \rangle  \otimes |\nu, l, m\rangle,
\end{align}
The total Hamiltonian for two atoms are their spin part and spacial motion parts, i.e.
$\hat{H}_{\mathrm{doub}} = \hat{H}_{\mathrm{spin}}
  + \hat{H}_{\mathrm{rel}}
  + \hat{V}$.
In the Eq.~\eqref{eq:basis} basis, the Hamiltonian can be written as matrix,
\begin{align}\nonumber
  H_{\mathrm{doub}} =& \frac{\hbar}{2}
  \begin{pmatrix}
    2\delta + \frac{2}{\hbar}E^{\mathrm{pseudo}}_{\nu, l} & \sqrt{2}\Omega_1 \eta & 0 \\
    \sqrt{2}\Omega_1 \eta &  \frac{2}{\hbar} E_{\nu, l} & \sqrt{2}\Omega_1 \\
    0 & \sqrt{2}\Omega_1 & -2\delta + \frac{2}{\hbar}E_{\nu, l}
  \end{pmatrix} \\
  =&\frac{\hbar}{2}
  \begin{pmatrix}
    2\delta + \frac{2}{\hbar}U_{\nu, l} & \sqrt{2}\Omega_1 \eta & 0 \\
    \sqrt{2}\Omega_1 \eta & 0 & \sqrt{2}\Omega_1 \\
    0 & \sqrt{2}\Omega_1 & -2\delta
  \end{pmatrix}
  + E^{\mathrm{pseudo}}_{\nu, l} I_3,
\end{align}
where $U_{\nu, l} =  E^{\mathrm{pseudo}}_{\nu, l} - E_{\nu, l}$ is the energy shift due to the pseudopotential interaction. $I_3$ is the $3\times 3$ identity matrix. The wave function overlap is defined as
\begin{align}
  \nonumber
  \eta =& \langle\nu, l, m |\nu, l, m\rangle_{\mathrm{pseudo}}
  = \int \mathrm{d}\mathbf{r} \, \Psi^{*}_{\nu, l, m}
  \Psi^{\mathrm{pseudo}}_{\nu, l, m} \\ \label{eq:overlap}
  =&  \int_0^{\infty} \mathrm{d}r  \,r^2 \psi^{*}_{\nu, l}
  \psi^{\mathrm{pseudo}}_{\nu, l}.
\end{align}
we should also choose the same finite cutoff rather than integrate from zero when we calculate this overlap integral.

When the RF pulse is in resonance with the states
${|T, \downarrow\downarrow \rangle \otimes |\nu, l, m\rangle}$
and
${|T, \uparrow\uparrow \rangle \otimes |\nu, l, m\rangle_{\mathrm{pseudo}}}$, i.e.
$\hbar\delta = \hbar\omega_{\mathrm{zs}} - \hbar\omega_{\mathrm{rf}} = - U_{\nu, l}/2$, the Hamiltonian can be written as
\begin{align}
  \label{eq:doub-hamiltonian}
  H_{\mathrm{doub}}
  =\frac{\hbar}{2}
  \begin{pmatrix}
    0 & \sqrt{2}\Omega_1 \eta & 0 \\
    \sqrt{2}\Omega_1 \eta & - U_{\nu, l}/\hbar & \sqrt{2}\Omega_1 \\
    0 & \sqrt{2}\Omega_1 & 0
  \end{pmatrix},
\end{align}
where we have dropped out some unimportant energy constant. They are the same as the formula in~\cite{Venu2023}. But thanks to the wave function solution for any partial wave, we can use Eq.~\eqref{eq:doub-hamiltonian} for any partial wave interacting cases.

\section{Results}\label{sec:results}

The two body states at time $t$, labeled as ${|\phi(t)\rangle}$ can be obtained by solving the time-dependent Schrodinger equation for the two particles Hamiltonian Eq.~\eqref{eq:doub-hamiltonian}.
The atom number in state ${|\uparrow\rangle}$, labeled as $N_{\uparrow}(t)$, can be observed experimently.
The Rabi oscillation for $N_{\uparrow}(t)$ at time $t$ are given in~\cite{Venu2023} as
\begin{widetext}
  \begin{align}
    \nonumber
    N_{\uparrow}(t) =&
    \left|\left(\langle T, \uparrow\downarrow| \otimes \langle \nu, l, m|\right)|\phi(t)\rangle\right|^2
    +
    2\left|\left(\langle T, \uparrow\uparrow |  \otimes \langle \nu, l, m|\right)|\phi(t)\rangle\right|^2 \\ \label{eq:rabi-ana}
    =& 2 \frac{\eta^2 \Omega_1^2}{(1+\eta^2)u^2} \left[
      \frac{1-\eta^2}{\eta^2} \sin^2\left( \frac{u}{2}t     \right)
    \right. \left.
      + \frac{8 u}{2 u - U_{\nu, l}/\hbar}\sin^2\left(
        \frac{2 u - U_{\nu, l}/\hbar}{8} t
       \right)
       + \frac{8 u}{2 u + U_{\nu, l}/\hbar}\sin^2\left(
        \frac{2 u + U_{\nu, l}/\hbar}{8} t
       \right)
     \right],
  \end{align}
\end{widetext}
where
\begin{align}
  u = \frac{1}{2} \sqrt{\left(\frac{U_{\nu, l}}{\hbar}\right)^2
   + 8(1+\eta^2)\Omega_1^2}.
\end{align}
We can see from Eq.~\eqref{eq:rabi-ana} that there are three oscillation frequencies in $N_{\uparrow}(t)$.
In the, $U_{\nu, l} \gg \hbar \Omega_1$, weak coupling limit, we can only consider the lowest frequency and neglect the other two frequencies. Thus the Rabi oscillation can be simplified as,
\begin{align} \label{eq:rabi-weak}
N_{\uparrow}(t)\approx
 \frac{8\eta^2}{(1+\eta^2)^2}\sin^2\left(
  \frac{2 u - U_{\nu, l}/\hbar}{8} t
 \right),
\end{align}
and the Rabi oscillation frequency is,
\begin{align}
  \Omega_{\uparrow} = \frac{2 u - U_{\nu, l}/\hbar}{4}.
\end{align}

\begin{figure}[htb]
  \centering
  \includegraphics{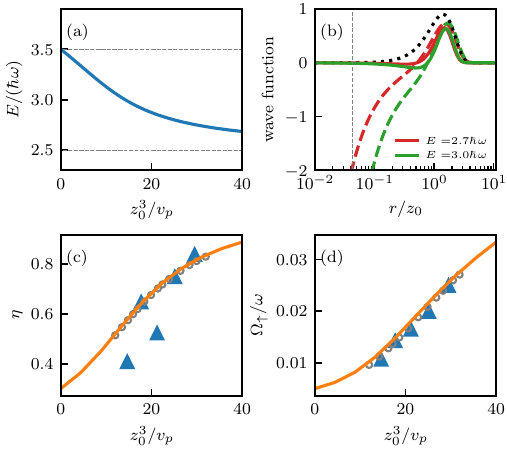}
  \caption{\label{fig:pair_osc_p}
  The Rabi oscillation for p-wave interacting $^{40}\mathrm{K}$ atoms in different interaction strengths. (a) The energy spectrum ($\nu=0$, $l=1$) we considered in the calculation.
  (b) The wave function and overlap. The dotted line is the free atom solution $r \psi_{\nu=0, l=1}$.
  The dashed line is interacting solutions $r \psi_{\nu=0, l=1}^{\mathrm{pseudo}}$.
  The solid line is their overlap $r^2 \psi_{\nu=0, l=1} \psi_{\nu=0, l=1}^{\mathrm{pseudo}}$.
  The grey dashed line indicates the cutoff $50~a_0$.
  (c) The wave function overlap $\eta$ as a function of scattering volume.
  (d) The oscillation frequency as a function of scattering volume.
  The grey circle and the solid triangle in (c) and (d) are the theoretic and experiment results in~\cite{Venu2023}.
  }
\end{figure}

We show the Rabi oscillation for p-wave interacting $^{40}\mathrm{K}$ atoms in Fig.~\ref{fig:rabi_osc}(a) both from Eq.~\eqref{eq:rabi-ana} and Eq.~\eqref{eq:rabi-weak}.
Here we use the parameters in experiment~\cite{Venu2023}, the trap frequency is $\omega = 2\pi \times 129~\mathrm{kHz}$.
The cutoff for the normalization factor and the wave function overlap was fixed as $50~a_0$, where $a_0$ is the Bohr radius.
The Rabi frequency of the RF pulse is $\Omega_1=2\pi\times 8.83~\mathrm{kHz}$. And take the RF frequency in resonance with the interaction energy in ${|T, \uparrow\uparrow \rangle \otimes |\nu, l, m\rangle_{\mathrm{pseudo}}}$ as $E = 3\hbar\omega$.
The atoms are prepared in state ${|T, \downarrow\downarrow \rangle  \otimes |\nu =0, l=1, m\rangle}$, then the RF pulse couples the atom to state ${|T, \uparrow\uparrow \rangle \otimes |\nu=0, l=1, m\rangle_{\mathrm{pseudo}}}$.
Thus we can get
\begin{align}
  \frac{U_{\nu, l}}{\hbar \omega} = 0.5 \gg \frac{\Omega_1}{\omega} \approx0.068,
\end{align}
which indicates it is in the weak coupling limit. As shown in Fig.~\ref{fig:rabi_osc}(a), the results from Eq.~\eqref{eq:rabi-ana} and Eq.~\eqref{eq:rabi-weak} agrees well.

Furthermore, we can see the three peaks more clearly by transforming the results of $N_{\uparrow}$ into the frequency domain.
We assume the oscillation is exponential decay with lifetime $3.4~\mathrm{ms}$, which is the typical lifetime of $^{40}\mathrm{K}$ atoms due to the two-body loss~\cite{Venu2023}. Then we can transform the Rabi oscillation into the frequency domain by Fourier transform, as shown in Fig.~\ref{fig:rabi_osc}(b). We can see that there are three peaks as we expected.

Next, we investigated the Rabi oscillation for different interaction strengths which are characterized by scattering phase shift.
For our concerned low energy scattering p-wave and d-wave interaction, the generalized scattering length is scattering volume $v_p$ and super volume $D_d$ correspondingly.
Experimentally, these two scattering parameters can be tuned by changing the magnetic field.

\begin{figure}[htb]
  \centering
  \includegraphics{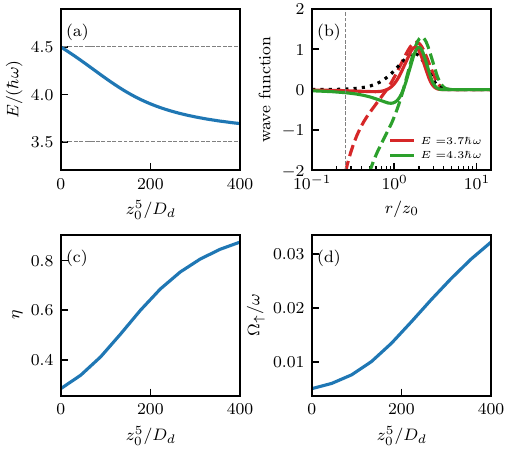}
  \caption{\label{fig:pair_osc_d}
    The Rabi oscillation for d-wave interacting $^{41}\mathrm{K}$ atoms in different interaction strengths. (a) The energy spectrum ($\nu=0$, $l=2$) we considered in the calculation.
  (b) The wave function and overlap. The dotted line is the free atom solution $r \psi_{\nu=0, l=2}$.
  The dashed line is interacting solutions $r \psi_{\nu=0, l=2}^{\mathrm{pseudo}}$.
  The solid line is their overlap $r^2 \psi_{\nu=0, l=2} \psi_{\nu=0, l=2}^{\mathrm{pseudo}}$.
  The grey dashed line indicates the cutoff $300~a_0$.
  (c) The wave function overlap $\eta$ as a function of super volume.
  (d) The oscillation frequency as a function of super volume.
  }
\end{figure}

The results for p-wave interacting $^{40}\mathrm{K}$ atoms are shown in Fig.~\ref{fig:pair_osc_p}.
Fig.~\ref{fig:pair_osc_p}~(a) shows the energy spectrum that we are concerned in the calculation. This branch is adiabatically connected to the $\nu = 0, l = 1$ mode, whose energy is $5 \hbar \omega/2$.
In Fig.~\ref{fig:pair_osc_p}~(b), we show the wave function both for two interacting and non-interacting particles in a harmonic trap. We chose two interaction strengths, with energy $E=2.7\hbar\omega$ and $E=3.0\hbar\omega$. The normalization factor are calculate with the cutoff $50~a_0$ indicated by the vertical grey dashed line.
In Fig.~\ref{fig:pair_osc_p}~(c) and (d), we repeated the theoretic and experiment results in~\cite{Venu2023} for wave function overlap $\eta$ and Rabi frequency $\Omega_{\uparrow}$, shown as the grey circle and triangle. We can see $\eta$ approach to unity when $v_p$ is small, this implies that the choice of the cutoff is proper.


The results for d-wave interacting $^{41}\mathrm{K}$ atoms are shown in Fig.~\ref{fig:pair_osc_d}. It is worth noting that the cutoff is $300~a_0$ which is much larger than the p-wave case. The choice of the cutoff should make the overlap $\eta$ close to unity in the non-interacting limit.

\section{Summary}\label{sec:summary}

High partial wave interacting quantum many-body systems attracted much interest in recent years, especially in ultracold atom experiments, which can realize such a system and the interaction between the atoms can be tuned by Feshbach resonance. However, the experiments are limited by three-body loss problems. In a recent experiment, the authors load the $^{40}\mathrm{K}$ ultracold atoms into a deep optical lattice and tune the magnetic to the p-wave Feshbach resonance.
They give the theoretic model for their p-wave case.
In this work, we combine the solution of interacting two-body problems in a harmonic trap and generalize this theoretic model to any partial wave cases.
Our results may be useful for further experiments and theoretic investigations.

\begin{acknowledgments}
  This work was supported by the National Key Research and Development Program of China (Grant No. 2022YFA1405301 and No. 2018YFA0306502), the National Natural Science Foundation of China (Grant No. 12022405 and No. 11774426), the Beijing Natural Science Foundation (Grant No. Z180013).
\end{acknowledgments}

\newpage
\bibliography{reference}

\end{document}